\documentclass[fleqn,usenatbib,useAMS]{mnras}

\usepackage{graphicx}	
\usepackage{amsmath}	
\usepackage{amssymb}	
\usepackage{multicol}        
\usepackage{bm}		
\usepackage{pdflscape}	
\usepackage{longtable}

\usepackage{hyperref}

\usepackage{natbib}
\setcitestyle{authoryear,open={(},close={)}}

\newcommand{\Rha}{$R\mathrm{_{h}}=0.6 \,\mathrm{pc}\,$}

\newcommand{\W}{$W\mathrm{_{0}}\,$}

\newcommand{\f}{$f\mathrm{_c}\,$}
\newcommand{\fb}{$f\mathrm{_c}=0.5\,$}
\newcommand{\fc}{$f\mathrm{_c}=1.0\,$}

\newcommand{\bse}{\textsc{bse}}
\newcommand{\bseS}{\textsc{bse} $\,$}
\newcommand{\bsepsn}{\textsc{bse-psn}}
\newcommand{\bsepsnS}{\textsc{bse-psn} $\,$}

\newcommand{\msun}{$\mathrm{M_{\odot}}$} 
\newcommand{\msunS }{$\mathrm{M_{\odot}}\,$} 

\usepackage[T1]{fontenc}
\usepackage{ae,aecompl}

\usepackage{newtxtext,newtxmath}

\title[Massive black hole formation in compact star clusters]{Black hole mergers in compact star clusters and massive black hole formation beyond the mass-gap}
\author[F. P. Rizzuto]{Francesco Paolo Rizzuto$^{1}$\thanks{Contact e-mail: \href{mailto:}{rizzuto@mpa-garching.mpg.de}},
Thorsten Naab$^{1}$, Rainer Spurzem$^{2,3,4}$\thanks{Research Fellow at Frankfurt Institute for Advanced Studies
(FIAS)}, Manuel Arca-Sedda$^{2}$ , \newauthor Mirek Giersz$^{5}$, Jeremiah Paul Ostriker$^{6,7}$, Sambaran Banerjee$^{8,9}$
\\
$^{1}$Max-Planck Institute for Astrophysics, Karl-Schwarzschild-Str. 1,D-85741, Garching, Germany
\\
$^{2}$Astronomisches Rechen-Institut, Zentrum für Astronomie, University of Heidelberg, Mönchhofstrasse 12-14,69120, Heidelberg, Germany
\\
$^{3}$National Astronomical Observatories and Key Laboratory of Computational Astrophysics, Chinese Academy of Sciences, 20A Datun Rd., 
\\
Chaoyang District, 100101, Beijing, China
\\
$^{4}$Kavli Institute for Astronomy and Astrophysics, Peking University, Yiheyuan Lu 5, Haidian Qu, 100871, Beijing, China
\\
$^{5}$Nicolaus Copernicus Astronomical Centre, Polish Academy of Sciences, ul. Bartycka 18, 00-716 Warsaw, Poland
\\
$^{6}$Department of Astronomy, Columbia University, New York, NY 10027, USA
\\
$^{7}$Department of Astrophysical Sciences, Princeton University, Princeton, NJ 08544, USA
\\
$^{8}$Helmholtz-Institut fur Strahlen- und Kernphysik (HISKP), Nussallee 14-16, D-53115 Bonn, Germany
\\
$^{9}$ Argelander Institut f\"ur Astronomie (AIfA), Auf dem H\"ugel 71, D-53121, Bonn, Germany
}

\date{\today}

\pubyear{2020}

\begin{document}
\label{firstpage}
\pagerange{\pageref{firstpage}--\pageref{lastpage}}
\maketitle

\begin{abstract}
We present direct N-body simulations, carried out with \textsc{nbody6++gpu}, of young and compact low metallicity star clusters with $1.1\times 10^5$ stars, a velocity dispersion of $\sim$ 10 $\mathrm{km/s^{-1}}$, a half mass radius $R_h=0.6$ pc, and a binary fraction of $10\%$ including updated evolution models for stellar winds and pair-instability supernovae (PISNe). Within the first tens of megayears of evolution, each cluster hosts several black hole (BH) merger events which nearly cover the complete mass range of primary and secondary BH masses for current LIGO/Virgo/Kagra gravitational wave detections. The importance of gravitational recoil is estimated statistically. We present several possible formation paths of massive BHs above the assumed lower PISNe mass-gap limit ($45 M_\odot$) into the intermediate-mass BH (IMBH) regime ($> 100 M_\odot$) which include collisions of stars and BHs as well as the direct collapse of stellar merger remnants with low mass cores. The stellar evolution updates result in the early formation of higher mass stellar BHs than for the previous model. The resulting higher collision rates with massive stars support the rapid formation of massive BHs. For models assuming a high accretion efficiency for star-BH mergers, we present a first-generation formation scenario for GW190521-like events – a merger of two BHs in the PISN mass-gap – which is dominated by star-BH mergers. This IMBH formation path is independent of gravitational recoil and therefore conceivable in dense stellar systems with low escape velocities. One simulated cluster even forms an IMBH binary (153$M_\odot$,173$M_\odot$) which is expected to merge within a Hubble time. 
\end{abstract}
\begin{keywords}

gravitational waves – methods: numerical – stars: black holes – stars: dynamics – stars: mass-loss – galaxies: star clusters: general.

\end{keywords}



\begingroup
\let\clearpage\relax
\endgroup
\newpage

\section{Introduction}

Before the first LIGO gravitational-wave (GW) detection, many theoretical models of stellar evolution predicted stellar black holes (BHs) masses to be lower than $30$ \msun. These models remained unchallenged for several years because all stellar BHs observed at the time had masses  $\lesssim 20$ \msunS \citep{Ziokowski2008,Ozel2010}.
Surprisingly, the first LIGO detection, GW150914, revealed components more massive than $30$ \msun \citep{Abbott2016}. Such masses had been predicted by stellar evolution models at low metallicity introducing a dependence between stellar winds mass loss and metallicity \citep[see][and references therein]{Woosley2002,Vink2001}. This highlights the importance of stellar evolution models for the correct interpretation/prediction of GW events. An accurate theory for the evolution of massive stars is particularly important to predict the mass distribution of stellar BHs at their formation. For this, precise models of stellar winds and a correct description of the last stages of the stellar evolution before the collapse are required. At the onset of stellar collapse, stars with sufficiently large helium cores undergo a phase of electron-positron pair production that in turn leads to one or more violent explosions. Depending on the initial mass of the core, the star can experience pulsation pair-instability supernovae (PPSN) getting partially destroyed or it can experience the more violent pair-instability supernovae (PSN) and is destroyed completely \citep{Fowler1964, Woosley2007, Woosley2017}. Due to (P)PSN, isolated massive stars are not supposed to collapse into BHs in the mass range of approximately $50 - 130$ \msun.
This gap in the stellar BH mass distribution is known as the (P)PSN mass gap. The mass limits of this gap are affected by various uncertainties and therefore they depend on the details of the stellar evolution adopted.
In this study, the assumed mass gap is $45 - 195$ \msun.

The 66 - 85 \msunS BH merger (GW190521) detected by the LIGO/Virgo/Kagra collaboration \citep{Abbott2020precession} has started a debate about the origin of the two objects involved in the collision as both of BHs fall in the (P)PSN mass gap. In particular, the mass of the primary in GW190521 ($85$ \msun) poses a challenge to theoretical models. In a recent paper, \cite{Vink2021} has proposed a modified stellar evolution model such that $\sim100$ \msunS stars lose only little mass via stellar winds, small enough for the stars to collapse into BHs in the range of $\sim 85$ \msun. A recent study, leveraging newly estimated uncertainties on (P)PSN mass loss, affirm that GW190521 could have formed through the classical isolated binary evolution \citep{Belczynski2020}. Other potential formation channels for BHs of such masses might involve
population III stars \citep{Liu2020, Kinugawa2021} or might depends on details of 
the rates of carbon-oxygen nuclear reactions \citep{Farmer2019}.

As both observed merging BHs in question are above the conventional mass gap lower limit of $\sim 50$ \msunS \citep{Farmer2019, Woosley2020}, and because there is evidence for residual eccentricity \citep{Romero2020, Gayathri2020} and/or spin precession \citep{Abbott2020precession}, many authors have argued that GW190521 could have been produced dynamically through repeated mergers of smaller BHs \citep{Abbott2020precession}. The most popular models involve multiple mergers of low mass BHs in dense and massive stellar environments with an escape velocity of a few hundred km/s \citep{ArcaSedda2019, Fragione2020, Romero2020, Kimball2020, DallAmico2021, Mapelli2021Hie}. According to these studies, young star clusters and globular clusters are less favoured because the recoil kick imparted to the merger remnant could eject the final product of a BH - BH merger and halt its mass growth too early.

However, many of these studies might have underestimated the importance of star - BHs collisions for the formation of GW events mass range of GW190521. In agreement with previous studies \citep{Giersz2015, DiCarlo2019, Kremer2020, Becker2020, Rizzuto2021, ArcaSedda2021}, the star cluster simulations we present in this paper show that low mass BHs can reach the (P)PSN mass gap or even the IMBH mass range through stellar mergers. A similar outcome could be produced invoking tidal capture events \citep{Stone2017}.
Moreover, they illustrate that GW190521-like events could be hosted by $\sim 10^5$ \msunS compact star clusters if the  BHs involved in the collisions also grow by the accretion of stellar material.

\bigbreak

This work is an extension of \citet{Rizzuto2021} where we have shown that a few hundred solar masses IMBHs can form in low metallicity compact star cluster environments via multiple collisions\footnote{As pointed out by Mark Gieles in private communication, the process that leads to the formation of massive objects in our simulation is not a runaway growth because the growth rate does not increase with growing mass.}. We use updated stellar evolution models with pulsation pair instability prescriptions and new stellar winds recipes \citep[following][]{Banerjee2020} to evolve the most promising initial conditions presented in \cite{Rizzuto2021}. We describe the details of the stellar evolution models in the next section (Section 2) and present the initial conditions in Section 3. In Section 4 we discuss our results in comparison to our previous work. In the final section, we discuss and summarise the main results.

\section{The Method}

 \begin{table*}
  \begin{tabular}{lccccccccccc}
    \hline
  Model Name       & $r\mathrm{_c}$ & $\rho_c$  & $R\mathrm{_h}\,$ & $\sigma$ & $t_{\mathrm{rh}}$ & $t_{\mathrm{s}}$  & \f   & IMBH   &  $M_{\mathrm{IMBH}}$ & $t_{\mathrm{form}}$  \\
        & [pc]    & [\msunS / pc $^3$]   & [pc] & [km/s]   &    [Myr] & [Myr]  & & &[\msun] & [Myr] \\
  \hline
  R06W61.0        & $0.19$ & $1.1\times10^5$   & $0.6$ & 15 & 56   &1.4& $1.0$ & $6/8$  & 102, 108, 153, 156, 172, 293   & 76, 3.9, 29, 21, 6.9, 44      \\
  R06W60.5        & $0.19$ & $1.1\times10^5$ & $0.6$ & 15 & 56   &1.4& $0.5$ & $4/8$  & 98, 99, 109, 113   &34,  12,  6.5,  4.3        \\                  
  \end{tabular}
  \caption{Model parameters of the cluster simulations: $r\mathrm{_c}\,$: initial core radius; $\rho\mathrm{_c}\,$: initial central density; \W: central potential parameter for the King density profile \citep{King1966}; $R\mathrm{_h}\,$: half mass radius; $\sigma$ : dispersion velocity;  $t_{\mathrm{rh}}$: half mass relaxation time; $t_{\mathrm{s}}$: segregation time scale for $100$ \msun; \f: fraction of mass absorbed by a compact object during a direct collision with a star; \# IMBH : Number of  BHs with masses $\gtrsim 100$ \msunS formed out of 8 realisations; $M_{\mathrm{IMBH}}$: IMBH masses; $t_{\mathrm{form}}\,$: IMBHs formation times.}
   \label{table:initial}
\end{table*}

\begin{figure}
    \includegraphics[width=\columnwidth]{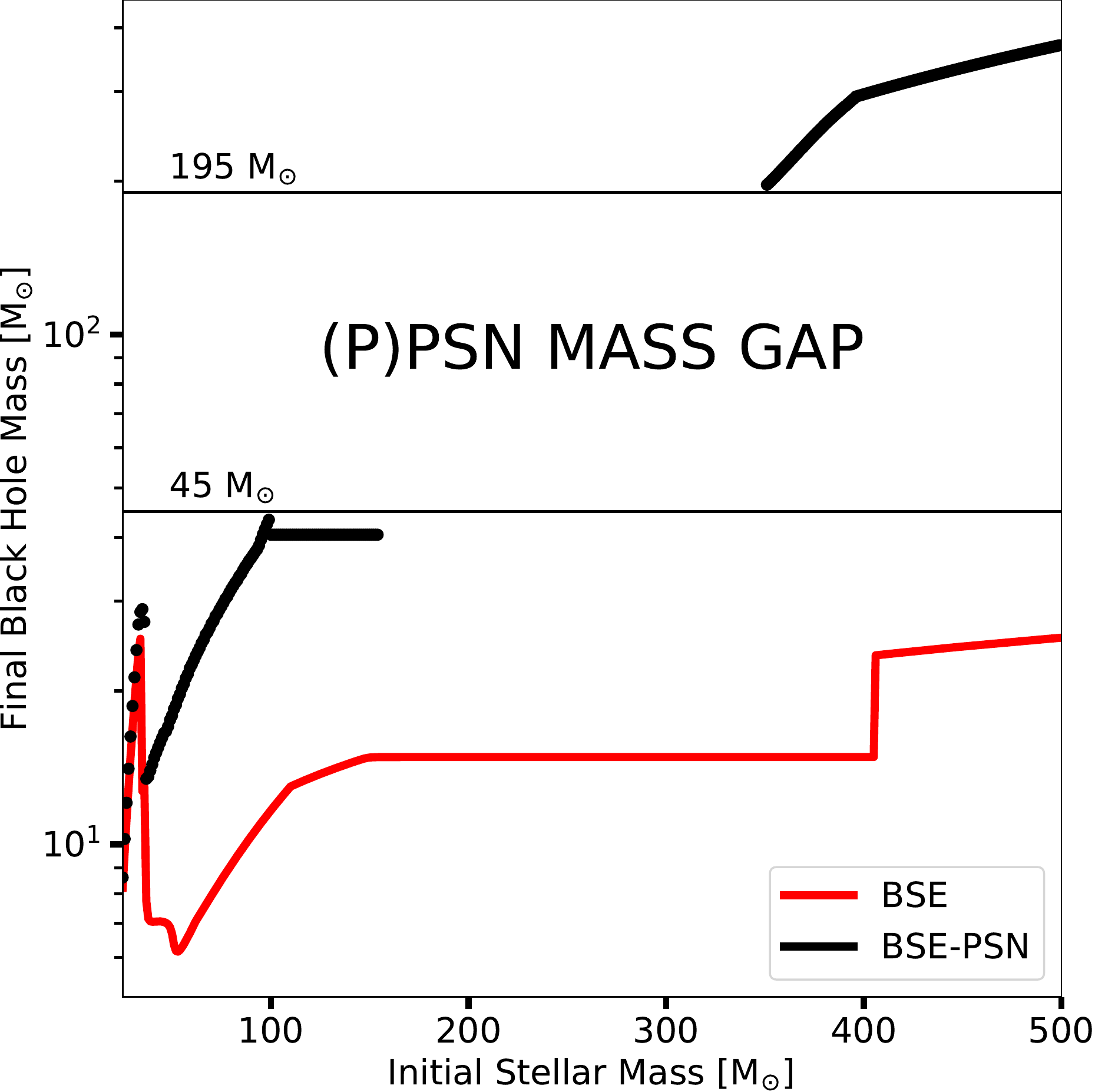}
    \caption{The final black hole mass as a function of initial stellar mass for stellar evolution model \bseS (red) used in \citet{Rizzuto2021} and the new model \bsepsnS (black) used here with updated stellar wind and pair instability supernova models. This is  for single stars in isolation at metallicity $Z = 0.0002$. For \bsepsnS the mass gap with with no stellar mass black hole remnant ranges from $45$ \msunS to $195$ \msun.}
    \label{fig:final_bh_mass}
\end{figure}

We present 16 new simulations of compact star clusters adopting the same initial conditions as in \cite{Rizzuto2021} using updated the stellar evolution models in the \textsc{nbody6++gpu} \citep{Wang2015} code. \textsc{nbody6++gpu} is capable of evolving star clusters with a realistic number of stars following the dynamical evolution of the system accurately as well as the stellar evolution of single and binary stars \citep[see also][]{2016MNRAS.458.1450W}. The integrator follows the dynamical evolution of binary stars, even in phases of mass loss or when one of the two stars experiences a supernova explosion. In this case, the new trajectories of the remnant and its companion are computed accurately. \citep[for more details see e.g.][and references therein]{Rizzuto2021}. In extension to \citet{Rizzuto2021}, the stellar evolution models of \textsc{nbody6++gpu} have been significantly updated and improved \citep[see][for more details]{Kamlah2021}. Most of the updates have been published for \textsc{nbody7} already in \cite{Banerjee2020}. The major updates are:
\begin{enumerate}
\item New stellar wind models following \cite{Belczynski2010}. which in turn follow the wind mass-loss rates given by \cite{Vink2001}.  With these models, BHs masses that originate from single stars depend strongly on the metallicity. For instance, a $100$ \msunS isolated main-sequence star would leave a  $15$ \msunS BH at solar metallicity ($Z = 0.02$). At very low metallicity ($Z = 0.0002$), however, it can form a BH of about $60$ \msun, in the absence of pair-instability models. 
\item Pair-instability supernova and pulsation pair-instability supernova models \citep[according to][]{Belczynski2016} incorporated in the remnant formation and supernovae models as described in \cite{Fryer2012}.  Stars with helium core with masses $> 40$ \msunS undergo a violent phase of mass loss. For helium cores in the range between $60 - 135 $ \msunS the star is completely destroyed.
\item New prescription of BHs and NSs natal kick velocities that explicitly depend on the fallback fraction \citep{Banerjee2020}. 
\item A model for electron capture supernovae (ECSN) following \cite{Podsiadlowski2004} and \cite{Gessner2018} that produces neutron stars with low-velocity kicks that are therefore likely retained in medium-size star clusters. \end{enumerate}

The gravitational energy loss and resulting merger of compact objects is computed following \citet{1963PhRv..131..435P} and the implementation is presented in  \citet{Rizzuto2021}. An important physical process that is not yet implemented in the code is relativistic recoil for compact object collisions. As we will see in the next sections, its absence might artificially enhance the probability of forming massive BHs \citep[see][for more details on relativistic recoil implementation in \textsc{nbody7}]{Banerjee2021}.

Hereafter, We use the term \bseS to refer to the previous stellar evolution prescription \citep{Rizzuto2021} while for the updated recipes we use the term \bsepsn. The updates in \bsepsnS have a strong impact on the mass distribution of the stars and their remnants in our simulations. With \bse,  an isolated massive star at metallicity $Z = 0.0002$, even as massive as $500$ \msun, can never collapse into stellar-mass BH more massive than $30$ \msun, as shown in Fig. \ref{fig:final_bh_mass} (red line). At the same metallicity, the updated model \bsepsnS with new stellar wind and the (P)PSN models, results in more massive stellar BHs. Main-sequence stars with masses in the range $80-100$ \msunS leave a remnant of about $40$ \msunS and very massive stars with masses of a few $\sim 10^2$ \msunS collapse directly into IMBHs, as illustrated by the black dots in Fig. \ref{fig:final_bh_mass}. The mass-gap, without any remnant black holes, is thus limited to $45 - 195$ \msunS (Fig. \ref{fig:final_bh_mass}) in our models.

\begin{figure}
    \includegraphics[width=0.9\columnwidth]{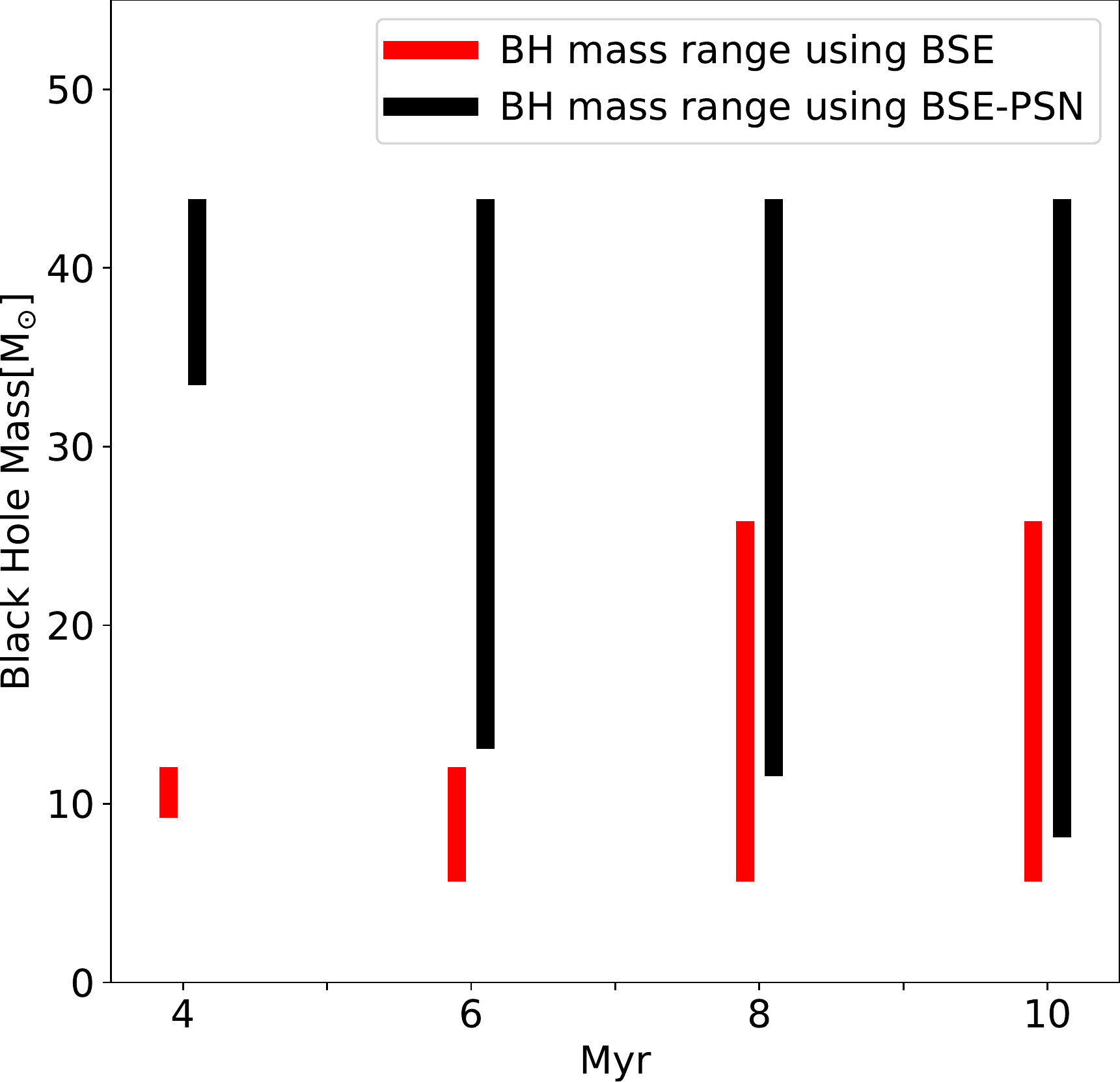}
    \caption{The plot shows in black (red) the BH mass range, at different times, produced by evolving heavy stars of $25 - 100$ \msunS in isolation using \bsepsnS(\bse). While  \bsepsnS produces BHs of $\sim$ 40 \msun, already after $4$ Myr,  \bseS starts to generate more massive BHs only after 6 Myr. Therefore, in the \bsepsnS (contrary to \bse) simulations, 40 \msunS BHs can coexist with massive stars of $\sim 50-70$ \msun.  The BHs have a higher probability to collide with more massive stars and grow more rapidly in mass.}
    \label{fig:BH_mass_range}
\end{figure}

The new stellar evolution model also affects the BH formation time. The black bars in Fig. \ref{fig:BH_mass_range} indicate the stellar BH mass ranges as a function of time. To estimate these mass ranges we evolved massive stars in isolation with \bsepsnS setting $Z = 0.0002$. Already after $4$ Myr BHs with $35 - 45$ \msunS form as remnants of the most massive stars ($\gtrsim 80$ \msun). Lighter stars explode later, pushing the lower BH mass limit down to $\sim 9$ \msunS after $10$ Myr. On the contrary, with the   \bseS prescriptions,  heavy stars in the range between $80$ \msunS and $100$ \msunS lose $ \sim 90 \%$ of their mass leading to early BHs with much lower masses of $\sim 9 - 12$ \msunS. With time the upper mass limit increases only to $\sim 25$ \msun  (red bars in Fig. \ref{fig:BH_mass_range}).
As we will see in Section \ref{sec:Results}, this difference between the old and the new stellar evolution models impacts the overall formation path of massive BHs: in the \bsepsnS simulations, stellar BHs above $35$ \msunS appear in the system when $\sim 70$ \msunS stars are still alive.  Therefore they have a higher probability to increase their mass by accretion of stellar material through close stellar interactions. 

There are also updates concerning the neutron star (NS) modelling. While in \bseS almost all NSs were ejected from the cluster due to the high natal kick velocities, in \bsepsnS the lower velocity kicks associated with ECSNe enable the formation of a NS population after ~ 10 Myr. 
However, for the short cluster evolution times considered in this study, none of our models shows any NS-NS or NS-BH merger.

Tidal capture events might play an important role in the growth of massive BHs. A recent analytical model, \cite{Stone2017} show that if every tidal capture event of a BH results in a collision, stellar-mass BHs, located in dense stellar environments, can grow up to $10^6$ \msunS within $\sim 1$ Gyr. Our simulations include tidal capture prescriptions that lead to rapid circularization rather than a collision. The radius of the final circular orbit is set assuming angular momentum conservation. Therefore, in general, tidal capture events do not lead to collisions and might not contribute directly to the BHs mass growth. 
Nevertheless, the circular BH-star  binary, once formed, have good chance to experience mass transfer or even mergers. 
Moreover, our instant tidal capture prescription requires knowledge of the orbital elements (pericenter and eccentricity) of the interacting objects. It is not activated during few-body chaotic encounters, where pericenter and eccentricity are not well-defined quantities. Therefore in some close interactions, we might underestimate the total number of tidal capture events.


\section{Initial Conditions}

\begin{figure*}
    \includegraphics[width=0.9\textwidth]{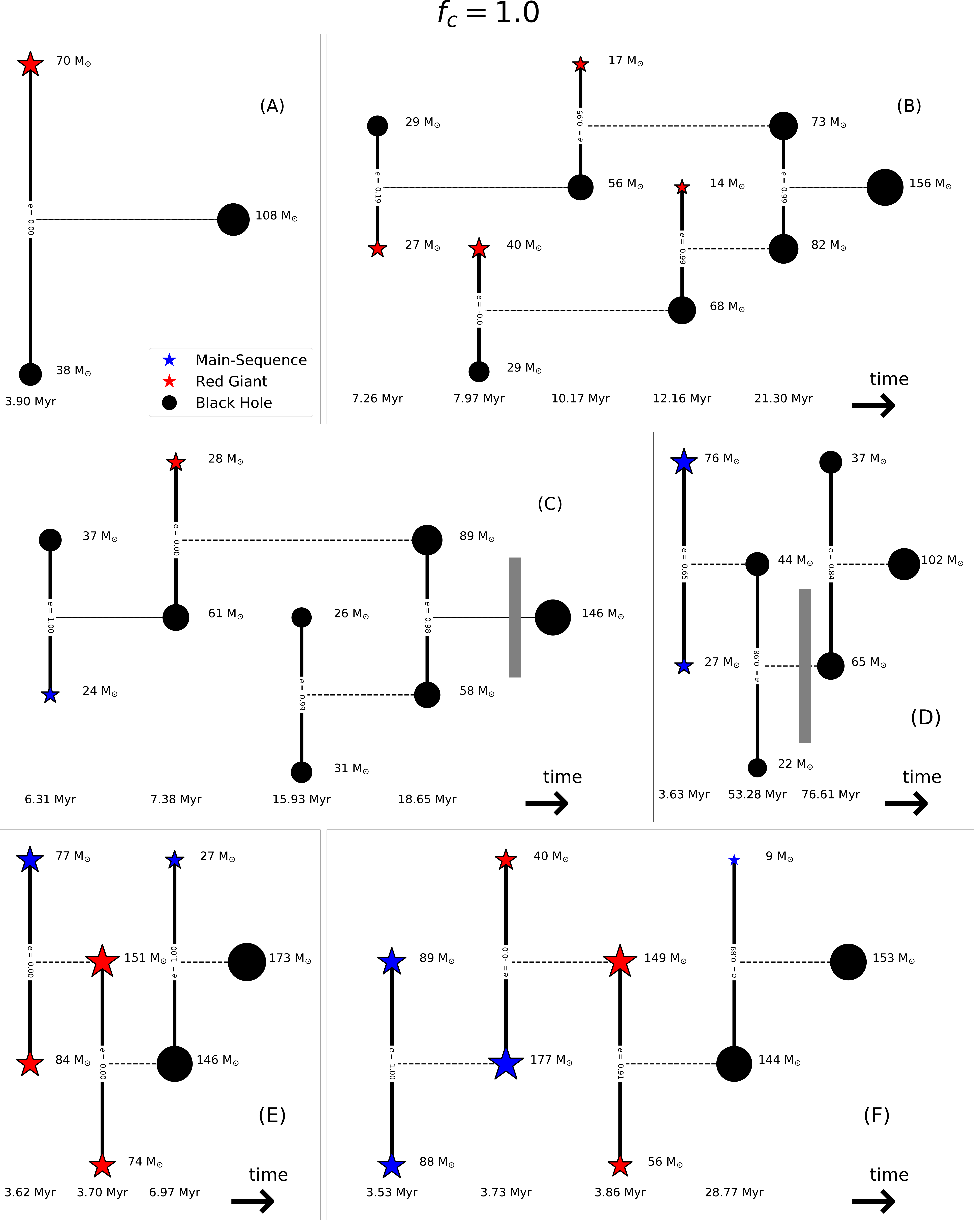}
    \caption{The formation paths of all six (A - F) BHs more massive than $100$ \msunS formed in the \fc model. Main sequence stars and red giants are indicated by blue and red stars, respectively. Black holes are shown by filled black circles. The symbol sizes indicate the respective masses. The collisions on the right of the grey vertical segment would be certainly suppressed by relativistic kicks. Time goes from left to right and we give the masses and the eccentricities at the time of the collision. Paths C and D involve BH mergers before a GW190521-like mass gap merger which could have been prevented by recoil kicks. Path B offers a  GW190521-like formation path only involving stellar accretion. Paths A, E, and F result in the formation of massive BHs in the IMBH range without stellar black hole mergers. For all examples, except C, we show the full formation path. The full path of C up to 100 Myr is shown in Fig. \ref{fig:collision_trees_most_massive_BH}.}
    \label{fig:collision_trees_fc1}
\end{figure*}

\begin{figure*}
    \includegraphics[width=0.9\textwidth]{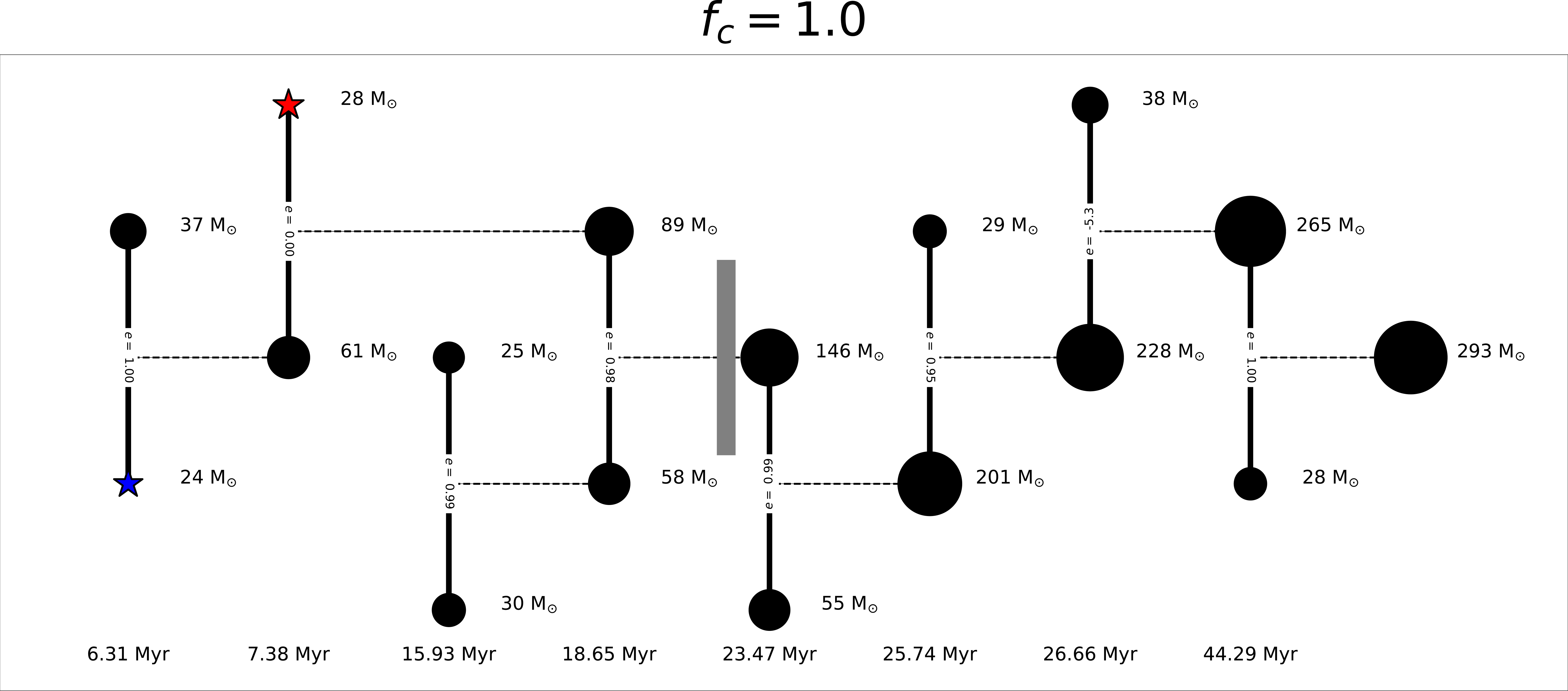}
    \caption{The formation path of the most massive BH formed in the 16 simulations. The final IMBH of 293 \msunS has formed through a chain of eight collisions. Panel C in Fig. \ref{fig:collision_trees_fc1} shows only the first 4. 6 collisions involve only BHs. This IMBH formed only because we did not include relativistic recoil kicks for BH - BH mergers. With kicks the mergers 
    after the grey vertical line would not have happened as one of the progenitors had left the system already.}
    \label{fig:collision_trees_most_massive_BH}
\end{figure*}

\begin{figure*}
    \includegraphics[width=0.9\textwidth]{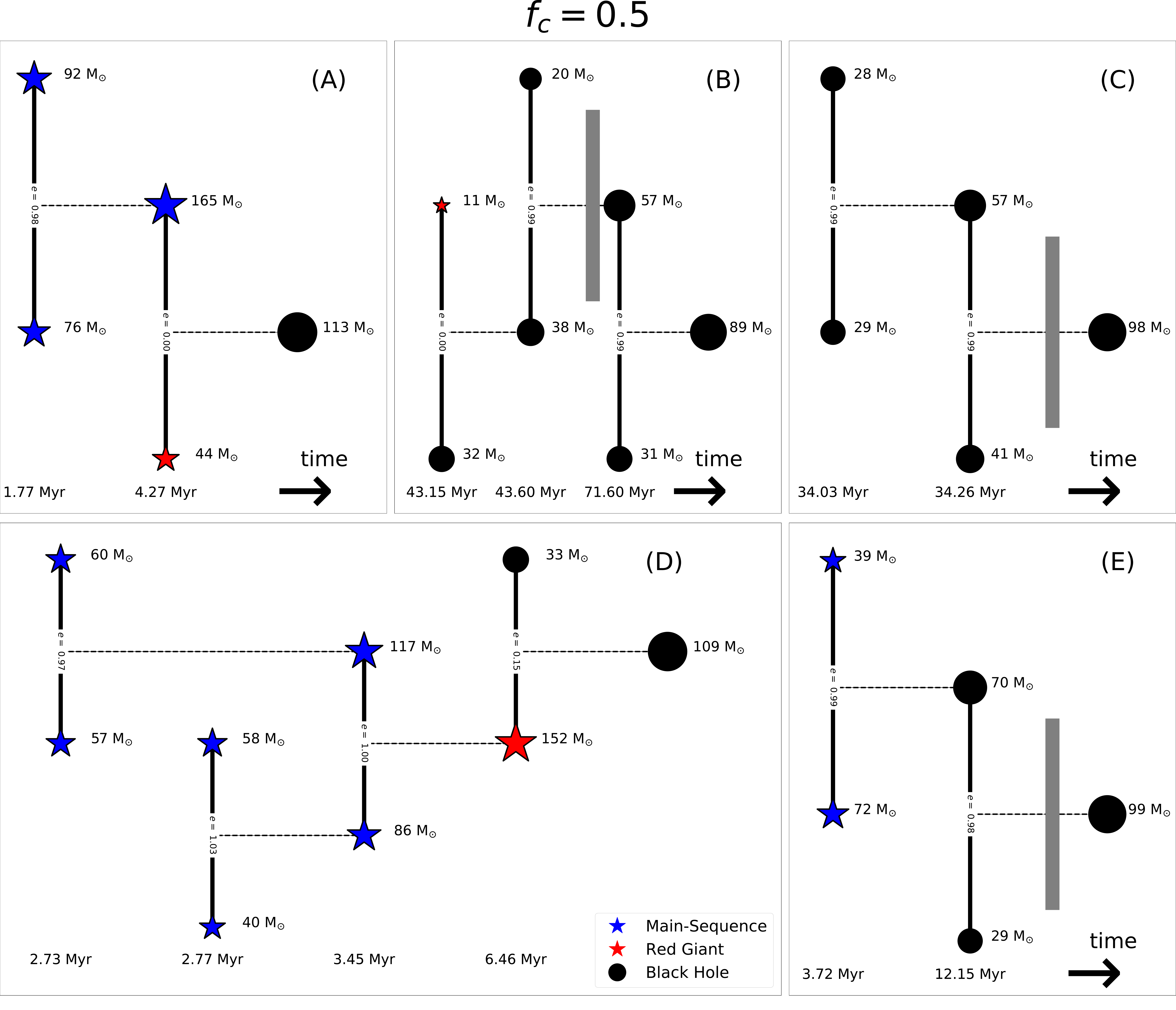}
    \caption{The formation paths of the five (A-E) massive black holes in the \fb model. The notation is the same as  in Fig. \ref{fig:collision_trees_fc1}. While this model also produces massive black holes, none of the formation paths involves a merger of two mass-gap BHs, like GW190521.}
    \label{fig:collision_trees_fc05}
\end{figure*}

Star clusters are expected to form from massive turbulent molecular clouds where gas is rapidly converted into stars. The high level of turbulence in these regions generates clumpy substructures in the whole star forming cloud. Numerical simulations indicate that violent relaxation erases initial substructure on a dynamical time scale \citep{Scally2002, Hurley2008} and leaves a spherical density profile compatible with an Elson, Fall and Freeman \citep{Elson1987} or a King \citep{King1966} density profile. Moreover, in the early stages,  gas and newly formed stars coexist in the same system. For moderately dense environments, only a small fraction of the cloud is converted into stars, the remaining gas is later ejected by stellar feedback and supernovae explosions.
Although star clusters at formation are very compact, since they typically have sizes $\lesssim 1$ pc \citep{Marks2012, Fujii2021}, gas ejection can lead to a rapid, strong expansion that can increase the size of the system to $\sim 5 - 10$ pc. This is especially true for low-mass star clusters ($10^3 - 10^4$ \msun).
These types of systems are expected to lose their primordial gas in an interval of time of about $\sim 9$ Myr \citep{Pelupessy2012, Fujii2021}. 

On the other hand,  observations show that young massive clusters (YMCs) such as NGC 3603, Trumpler 14, and Arches are devoid of dense gas despite being about $\sim 2$ Myr old \citep[sse][and citations therein]{Longmore2014}.
All these systems are very compact, with a photo-metric mass $> 10^4$ \msunS and and radii respectively equal to $0.7, 0.5, 0.4$ pc \citep{Zwart2010}. They are likely to originate from very dense gaseous environments, which thanks to their high density can maintain a high stellar formation efficiency, rapidly converting most of the gas into stars. In other words, these  YMCs remove a big fraction of their primordial gas through exhaustion rather than ejection \citep{Longmore2014}. Therefore they can retain their initial compactness. The stellar formation phase for these systems has a very short duration, leaving a star cluster with a very narrow age spread \citep[see e.g.][for the simulated formation of such systems in a galactic environment]{2019ApJ...879L..18L,2020ApJ...891....2L}.

A recent three-dimensional radiation hydrodynamical simulations of YMCs formation support the picture just presented, indicating that stellar feedback suppresses star formation only in clouds with low surface density; When the initial surface density exceeds a certain threshold most of the gas is converted into stars \citep{Fukushima2021}. In this last case, the star formation phase can last as little as 1 Myr.

Therefore, compact YMCs are ideal systems to be followed with direct N-body simulations, in spite of the fact that they are little affected by processes associated with gas ejections. We adopted the same initial conditions of the eight realizations of the model R06W6 presented in \cite{Rizzuto2021}. The model represents the state of young compact star clusters of $110.000$ stars with $10\%$ in binaries after gas removal. Their initial densities follow a King profile with initial half mass radius \Rha and central potential parameter \W\footnote{The central potential parameter is defined as \W$=\frac{\psi(0)}{\sigma^2}$  where $\psi(0)$ is the potential at the centre of the cluster and $\sigma^2$is a parameter linked to the velocity dispersion of the system. See \citep{King1966} for more details.} $=6$. Again, we assume that the systems maintain their initial compactness as the initial gas is removed by exhaustion from star formation. In addition, we assume the star formation phase is short-lived \citep[see e.g][]{2019ApJ...879L..18L}. Hence we neglect the initial age spread and we initialise zero-age main-sequence stars with a Kroupa initial mass function \citep{kroupa2001}. All the systems started with a primordial binary fraction of $10\%$,  initialised with a uniform semi-major axis distribution on a logarithmic scale from $0.001$ AU to $100$ AU, a uniform distribution of mass ratios, and thermal distribution of eccentricities. 

In the simulations we allow BHs to accrete mass from a star in case of a collision. The fraction of stellar mass absorbed by a BH during such a star - BH collision, which we define as collision factor \f,  has a significant impact on the formation of IMBHs \citep[see][]{Rizzuto2021}. This mass fraction, in general, is difficult to estimate, as it depends on the relative velocity of the two colliding objects as well as the internal structure of the stars, etc. A quantitative assessment of this number requires complex hydrodynamical simulations. In this work, we treat this fraction \f as a free parameter. We run  $8$ simulations setting \fb, and the other $8$ with \fc.

\section{Results} \label{sec:Results}

 In Fig. \ref{fig:collision_trees_fc1}, Fig. \ref{fig:collision_trees_most_massive_BH}, and Fig. \ref{fig:collision_trees_fc05} we show the formation paths of the most massive BHs in the simulations for \fc and \fb, respectively. Overall, these  simulations with updated binary stellar evolution models confirm our previous result \citep{Rizzuto2021} that massive black holes in the mass-gap and even IMBHs more massive than $100$ \msunS can form in compact star clusters through multiple mergers. With the new stellar
prescriptions, the runs registered an even higher probability to form
massive BHs. The latter build up their mass through at least one of
these three channels:
\begin{enumerate}
\item  Star - BH mergers: stellar BHs increase their mass by collisions with one ore more massive stars (see for instance panels A and B of Fig. \ref{fig:collision_trees_fc1} and panel D of Fig. \ref{fig:collision_trees_fc05}). 
\item BH - BH mergers: BHs grow by mergers with other BHs (i.e. panel C of Fig. \ref{fig:collision_trees_fc1} and panel C of Fig. \ref{fig:collision_trees_fc05}).
\item The collapse of stars that are the product of previous collisions: a VMS formed through a series of subsequent collisions could acquire a large hydrogen-rich envelope maintaining a small helium core ($M_c < 45$ \msun), it is therefore little or not affected by (P)PSN  and can collapse directly into an IMBH (i.e. panel F of Fig. \ref{fig:collision_trees_fc1} and panel A of Fig. \ref{fig:collision_trees_fc05}.
\end{enumerate}
This last formation channel was also observed in a recent set of N-body simulations by \citet{DiCarlo2021}, who also assume no mass loss in stellar merger events. The importance of stellar mergers, however, depends on the prescription used to model collisions between stars. Our assumption of no mass loss might overestimate the impact of stellar mergers. For example, \cite{Glebbeek2009} use a detailed stellar evolution code and take into account the internal structure of massive stars formed through multiple collisions. They indicate that steller mergers likely lose most of their mass in the form of stellar winds. Moreover, our method does not account for remnant rotation.  Even though the consequences of rotation have not yet been completely explored \citep{Burrows2021}, high stellar spin is expected to enhance the mass-loss rate \citep{Maeder2000}.

In general, at least two of the three processes mentioned above are always involved in the formation of a massive BH. For instance, the IMBH shown in panel C of Fig. \ref{fig:collision_trees_fc1} forms through the first and the second channel; while panel E of Fig. \ref{fig:collision_trees_fc05} presents an IMBH generated via the second and the third channel.  The most massive BH in all simulations has a mass of $293$ \msun. This object forms through $8$ collisions $6$ of them are BH - BH collisions as shown in Fig. \ref{fig:collision_trees_most_massive_BH}. This long chain of BH mergers happens when, as in our simulations, relativistic recoil kicks are not considered. When two BHs collide, the relativistic recoil kick velocity depends on the mass ratio as well as the spin amplitude and direction of the two colliding objects. It can reach a few thousand kilometres per second \citep{Campanelli2007, Baker2008,Lousto2009, Kulier2015, Morawski2018, Zivancev2020}. Since the typical escape velocity in our clusters is ~20-40km/s, it seems extremely unlikely that the $293$ \msunS IMBH with this formation path can form in star cluster environments. 

The first four collisions on the chain of mergers that lead to the $293$ \msunS are reported in panel (C) of Fig. \ref{fig:collision_trees_fc1}. Overall this panel shows two BH - star mergers and two BH - BH collisions that in the end generate a $146$ \msunS IMBH. Since we expect that most stellar-mass BHs are born with a low initial spin, as indicated partially by the current GW data \citep{Abbott2019, Miller2020, Roulet2020},
and suggested by recent theoretical models \citep{Fuller2019},
the $146$ \msunS IMBH has a non-negligible probability to form even with the inclusion of relativistic recoil as we shall discuss below. 

Therefore, the \fc simulations provide two possible formation paths for GW190521. Path C is a second-generation event and has a low probability due to a first generation BH merger, as discussed above. Path B is more likely as the BH merger event itself is a first generation BH merger.

\subsection{The impact of  stellar evolution updates}

\begin{figure*}
    \includegraphics[width=0.9\textwidth]{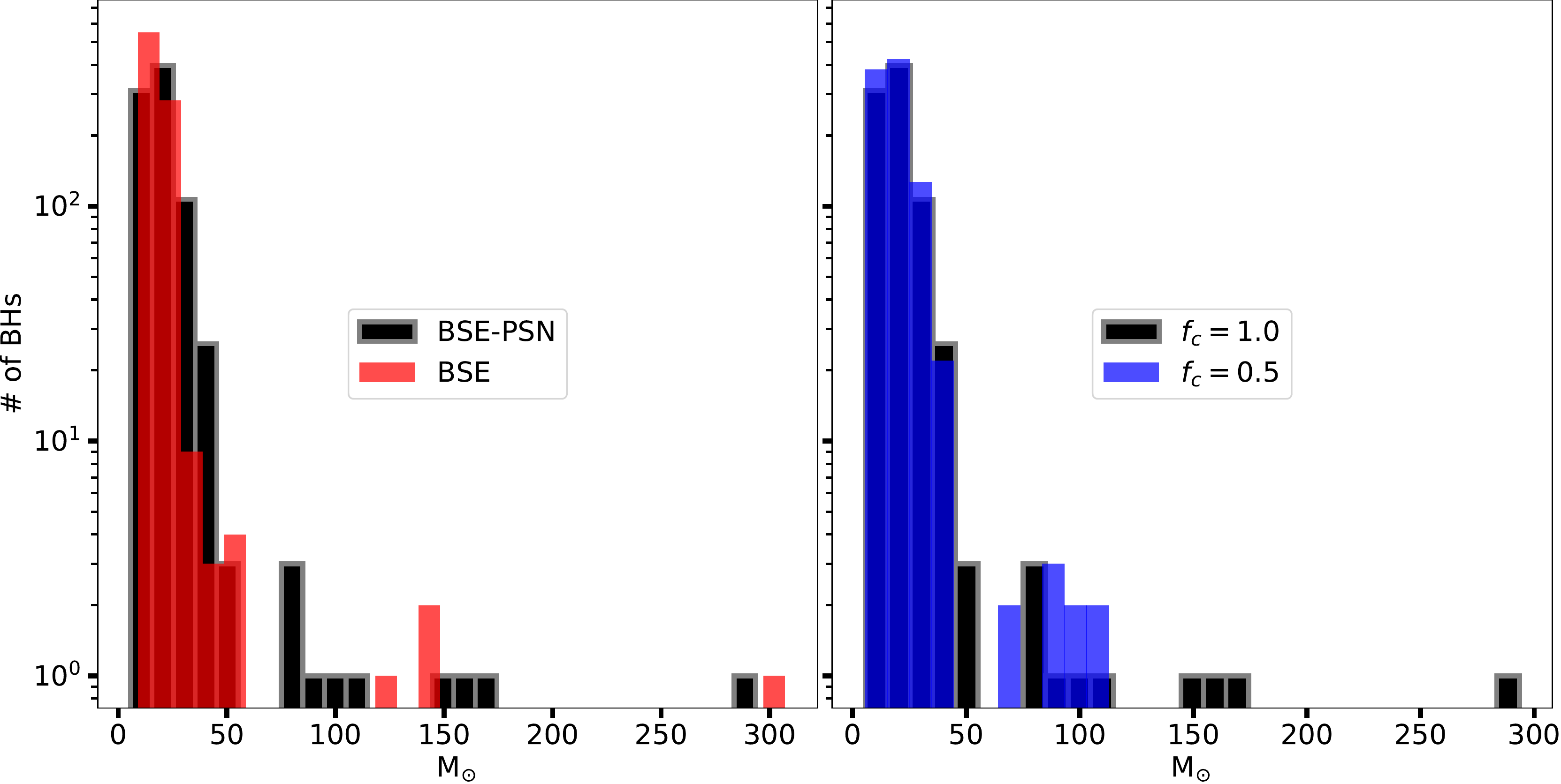}
    \caption{{\it Left panel:} Mass distribution of all BHs formed in the \bseS model (red) and the \bsepsnS model (black) with \fc after $100$ Myr of evolution. Simulations with \bsepsnS form more and more massive BHs in the (P)PSN mass gap and the IMBH mass regime. {\it Right panel:} Comparison of the BH mass distribution of the \bsepsnS model with \fc (black, same as left panel) to the model with \fb (blue). Lower accretion fractions for star-BH collisions result in less massive BHs.}
    \label{fig:bh_distribution_all}
\end{figure*}

\begin{figure*}
    \includegraphics[width=0.9\textwidth]{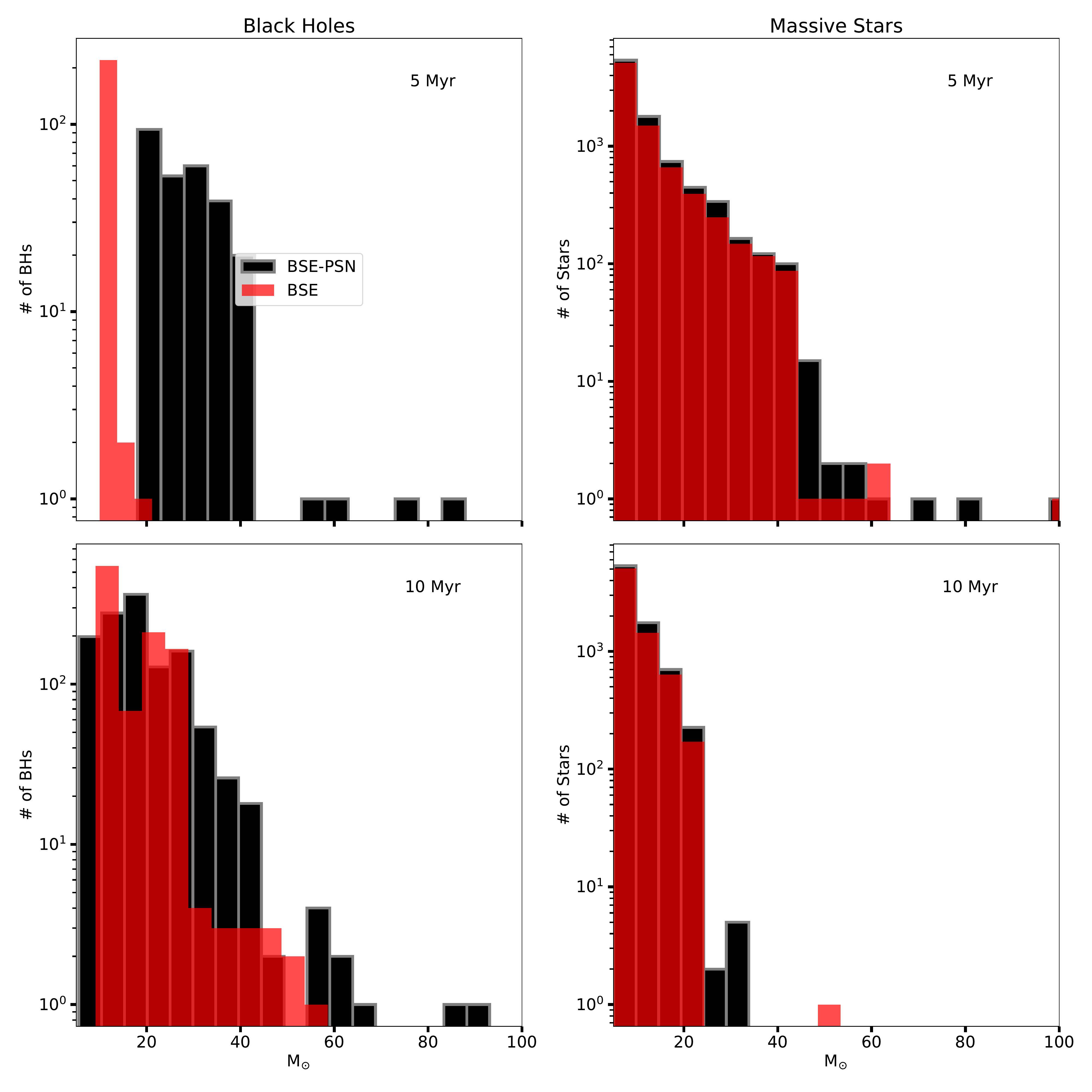}
    \caption{The two left (right) panels show the BH (massive stars) mass distribution after 5 Myr (top) and 10 Myr (bottom) of cluster evolution. In the \bsepsnS simulations (black histograms) $40$ \msunS BHs form early in the simulation while massive stars $\sim 60$ \msunS are still present (top panels). In contrast, the \bseS clusters contains only $\sim 10$ \msunS BHs after  $5$ Myr (red histograms). After 10 Myr the \bsepsnS model formed more massive BHs (bottom left panel) with a comparable stellar mass distribution (bottom right panel).}
    \label{fig:bh_distribution_old_vs_new}
\end{figure*}

After 100 Myr of cluster evolution, the BH mass distribution of the updated \bsepsnS model is very similar to the previous results as shown in the left panel of Fig. \ref{fig:bh_distribution_all}). The typical formation paths for the most massive BHS, however, differ significantly. In the \cite{Rizzuto2021} simulations (red histogram of Fig. \ref{fig:bh_distribution_all}, left panel) the BHs in the IMBH mass regime manly through mergers of BHs with VMSs: massive star experiences multiple stellar collisions reaching masses of $200-400$ \msun. A large fraction of this is then accreted by a stellar BH in a collision. Typically, it is a single collision between a stellar BH and a VMS that determines whether or not the cluster forms an IMBH. While this formation path can also occur in \bsepsnS (see e.g. panel D of Fig. \ref{fig:collision_trees_fc05}) it is not the dominant one. 

In the \bsepsnS simulations, stellar BHs, before reaching the IMBH mass range, typically experience multiple collisions with massive stars and other BHs. In several cases, IMBHs form entirely, or almost entirely, via repeated BH mergers, as illustrated in panels B and C of Fig. \ref{fig:collision_trees_fc05} and panel D of Fig. \ref{fig:collision_trees_fc1}. These formation paths are extremely rare in \bse; only 1 out of 80 simulations formed an IMBH through multiple BH - BH collisions.

The origin of this difference is the collision rate. The updated \bsepsnS model has about four times higher collision rate between massive stars and stellar BHs and three times higher BH-BH collision rate than \bseS, and BH - BH.  To understand this difference we briefly review the early cluster evolution. Within the first 4 Myr, the clusters undergo core-collapse driven by mass segregation of the most massive objects. Right after this initial collapse fast expansion is triggered by the mass loss of massive stars. For \bsepsnS massive stellar BHs form rapidly (see Fig. \ref{fig:BH_mass_range} and the top left panel of Fig. \ref{fig:bh_distribution_old_vs_new}) and already populate the inner part of the cluster while the central density is at its maximum and when massive stars with a large collision cross-sections are still alive (top right panel of Fig. \ref{fig:bh_distribution_old_vs_new}). High collision rates for star-BH and BH-BH mergers are the results.

This is not the case for \bseS simulations. The first BHs that appear in the core-collapse phase of the \bseS simulations have a mass of only $\sim 10$ \msunS (see top left panel of Fig. \ref{fig:bh_distribution_old_vs_new}) and they need a relatively long time to reach the inner part of the system, they typically reach the centre. At this time the clusters are not very dense anymore and $\gtrsim 40$ \msunS stars are already gone. More massive stellar BHs ($\sim 30$ \msun) start to populate the \bseS simulations only after the most massive stars have already exploded (bottom panels of Fig. \ref{fig:bh_distribution_old_vs_new}). 

In contrast, $\sim 40$ \msunS BHs populate the \bsepsnS clusters while massive stars are still alive (top panels of Fig. \ref{fig:bh_distribution_old_vs_new}). Due to their short segregation time, these BHs can reach the core of the cluster, while it is still dense, and collide with one or more massive stars. At the same time while sinking into the centre they support and enhance the core collapse.
To summarize, in the \bseS clusters, since the first stellar BHs have low mass (10 \msun), they have little chance to reach the centre when it is still very dense; Instead, in \bsepsnS the first stellar BHs that appear in the system are more massive (40 \msun), therefore, they can reach the centre while it is still compact and therefore they have a higher probability to collide with other objects.

Overall, the simulations with \bsepsnS seem to generate a larger number of massive BHs. The left panel of Fig. \ref{fig:bh_distribution_all} shows that \bsepsnS formed about 10 BHs more massive than 75 \msun, while \bseS simulation only form 4 BHs. This difference is also connected with the fact that \bsepsnS can form massive BHs, as we mentioned previously, through direct collisions of VMSs. The direct collapse channel cannot occur in \bseS because the stellar winds prescriptions are extreme and all VMSs lose most of their material before the collapse.

\subsection{Comparison between \fb and \fc}

\begin{figure*}
    \includegraphics[width=\textwidth]{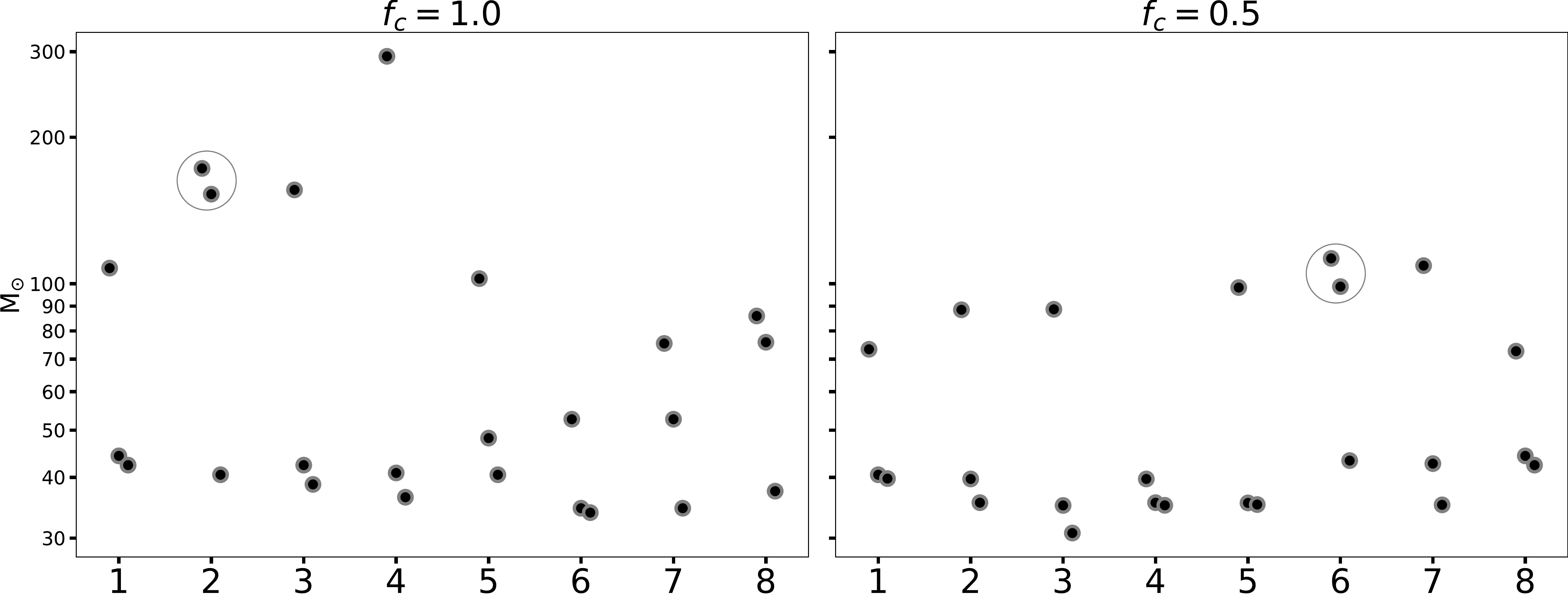}
    \caption{The mass of the three most massive BHs in each of the 8 realisations of the cluster simulations with \fc (left panel) and \fb (right panel). The first massive black hole is typically heavier in the \fc simulations, while the second and third have comparable masses to \fb models. }
    \label{fig:3_most_massive_bhs}
\end{figure*}

\begin{figure}
    \includegraphics[width=0.8\columnwidth]{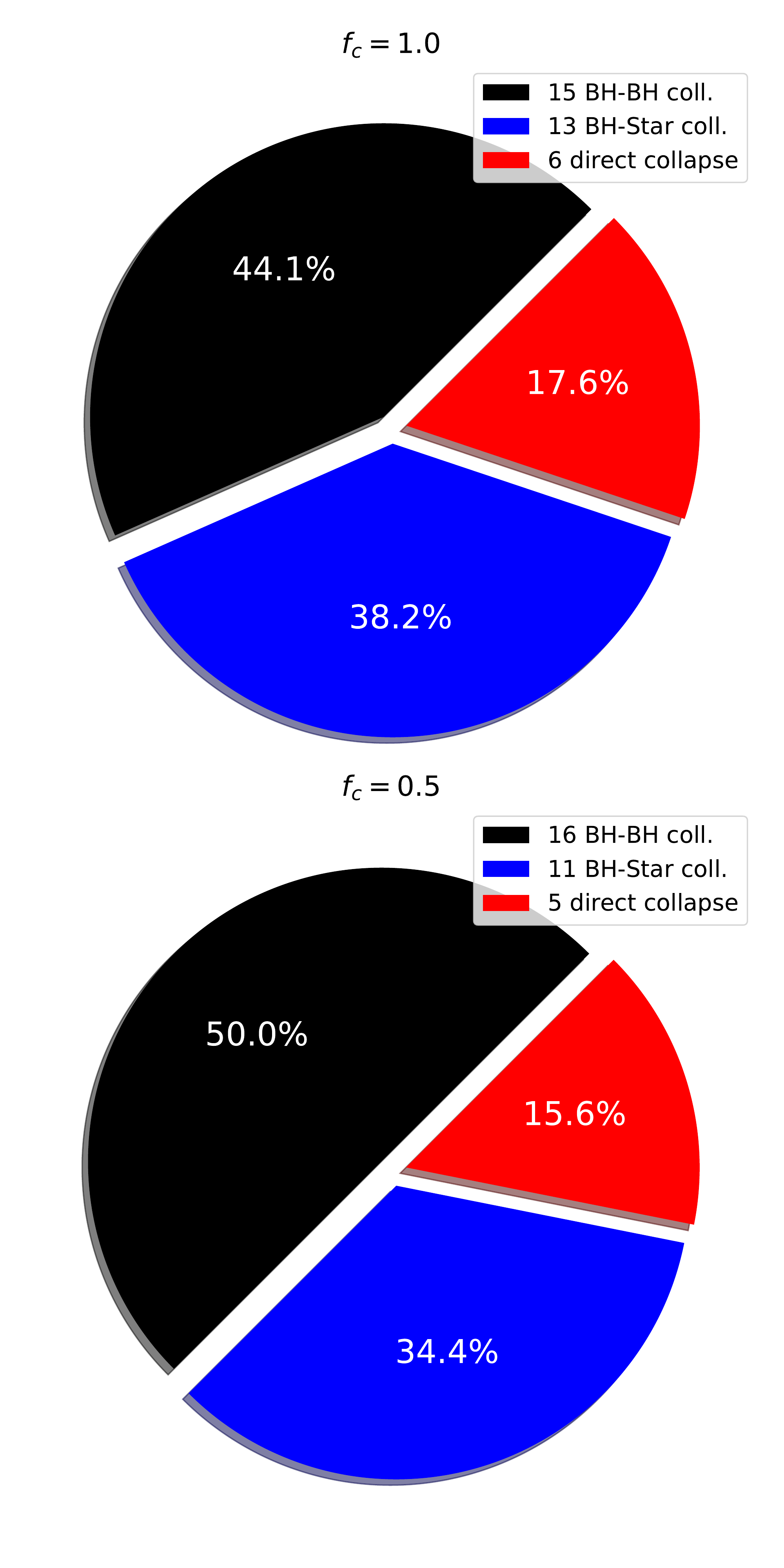}
    \caption{The Fig. shows in what percentage BH - BH mergers, BH - star coalescence, and direct collapse of stars (produced by previous collisions) contribute to the re-population of the (P)PSN mass gap in our simulations.}
    \label{fig:populate_mass_gap}
\end{figure}

Fig. \ref{fig:3_most_massive_bhs} compares the most massive BHs formed in the \fc simulations with the \fb simulations. As expected the former is more likely to generate BHs with higher masses. \fc models appear to be two times more likely to form IMBHs since already a single star - BH collision is enough to generate a BH heavier than $100$ \msunS (as shown in panel (A) of Fig. \ref{fig:collision_trees_fc1}). This can also be seen in the right panel of Fig. \ref{fig:bh_distribution_all} showing the BH mass distributions generated by the two models: below $\sim 40$ \msunS the two distributions are very similar. However, only the \fc model results in the formation of IMBHs above 150. The massive BHs of \fb model are lower masses in the range between $60$ and $110$ \msun.

As we mentioned earlier, because of the absence of observational and theoretical constraints, we choose \f to be a free parameter. The constraints on \f  presented in the literature refer to tidal disruption events between a low mass star and a supermassive BH. They estimate an \f in the range between  $\sim 0.2-0.5$ \citep{Shiokawa2015, Lu2020, Wen2020} because most of the mass is lost in high-speed jets formed during mass accretion. 
However, when a low mass BH collides with a massive star, the scenario might be very different.
Massive stars form very rapidly a compact core, so it might not be surprising that the actual core of the star is entirely absorbed by the BH right after the collision.
Later on, the resulting BH, immersed in the envelope of the star, might continue to absorb stellar material.
Due to the large size of the envelope of a massive star, the high-speed jets might be contained by the envelope itself, keeping most of the gas in the vicinity of the black hole. This situation resemble a similar scenario as described in \cite{Safarzadeh2020} and \citet{Rice2021}. Under these assumptions, \f might be close to unity.

The charts in Fig. \ref{fig:populate_mass_gap} visualise the formation paths of the BH more massive lower limit of the (P)PSN mass gap ($> 45$ \msun) for the \fc and \fb models. Most stellar BHs reach the mass gap by a merger with another BH. The next most important path is a collision with a star, followed by the direct collapse of a stellar merger remnant.  The relative importance of the formation paths does not change significantly with the assumed accretion fraction - even setting \fb, the star - BH collision channel produced $11$ BHs in the mass gap just $2$ less than the one produced with \fc.

 \begin{table}
  \begin{tabular}{lccccc}
      \hline
  BH merger    & $V_{\mathrm{esc}}$ (km/s)  &   P$_1$   & P$_2$ & P$_3$ \\
  \hline
  Fig. \ref{fig:collision_trees_fc1} panel B Gen-a   & 38.1     & 3.05 \%   &   82.2  \%   &   100.0  \%     \\
  Fig. \ref{fig:collision_trees_fc1} panel C Gen-a   & 40.3     & 1.91 \%   &   0.00  \%  &   0.00  \%       \\
  Fig. \ref{fig:collision_trees_fc1} panel C Gen-b   & 38.6     & 0.00  \%  &   0.00  \%  &   0.00  \%      \\
  Fig. \ref{fig:collision_trees_fc1} panel D Gen-a   & 29.2      & 0.00 \%  &   0.00  \%  &   0.00  \%      \\
  Fig. \ref{fig:collision_trees_fc1} panel D Gen-b   & 25.8     & 0.00  \%  &   0.00  \%  &   0.00  \%      \\
  Fig. \ref{fig:collision_trees_fc05} panel B Gen-a  & 31.2     & 0.00  \%  &   0.00  \%  &   0.00  \%      \\
  Fig. \ref{fig:collision_trees_fc05} panel B Gen-b  & 27.1     & 0.00  \%  &   0.00  \%  &   0.00  \%      \\
  Fig. \ref{fig:collision_trees_fc05} panel C Gen-a  & 35.7     & 4.44  \%  &   99.0  \%  &   100.0  \%    \\
  Fig. \ref{fig:collision_trees_fc05} panel E Gen-a  & 41.5     & 0.00   \%  &   0.00  \% &   0.00  \%      \\
 \hline
  \end{tabular}
  \caption{The table reports, for each BH - BH collision illustrated in Fig. \ref{fig:collision_trees_fc1} and \ref{fig:collision_trees_fc05}, the retention probability of the BH merger remnant in the star cluster. $V_{\mathrm{esc}}$: cluster escape velocity of the BH merger remnant. P$_1$: retention probability for an initial Gaussian spin distribution peaked on $S_1 = 0.2$, with a $\sigma = 0.2$ for the stellar BHs.  P$_2$: retention probability  for $S_2 = 10^{-2}$. P$_3$: retention probability for $S_3 = 10^{-3}$.}
 \label{table:spins}
 \end{table}

\subsection{Comparison with LIGO/Virgo/Kagra gravitational wave detections}

\begin{figure*}
    \includegraphics[width=\textwidth]{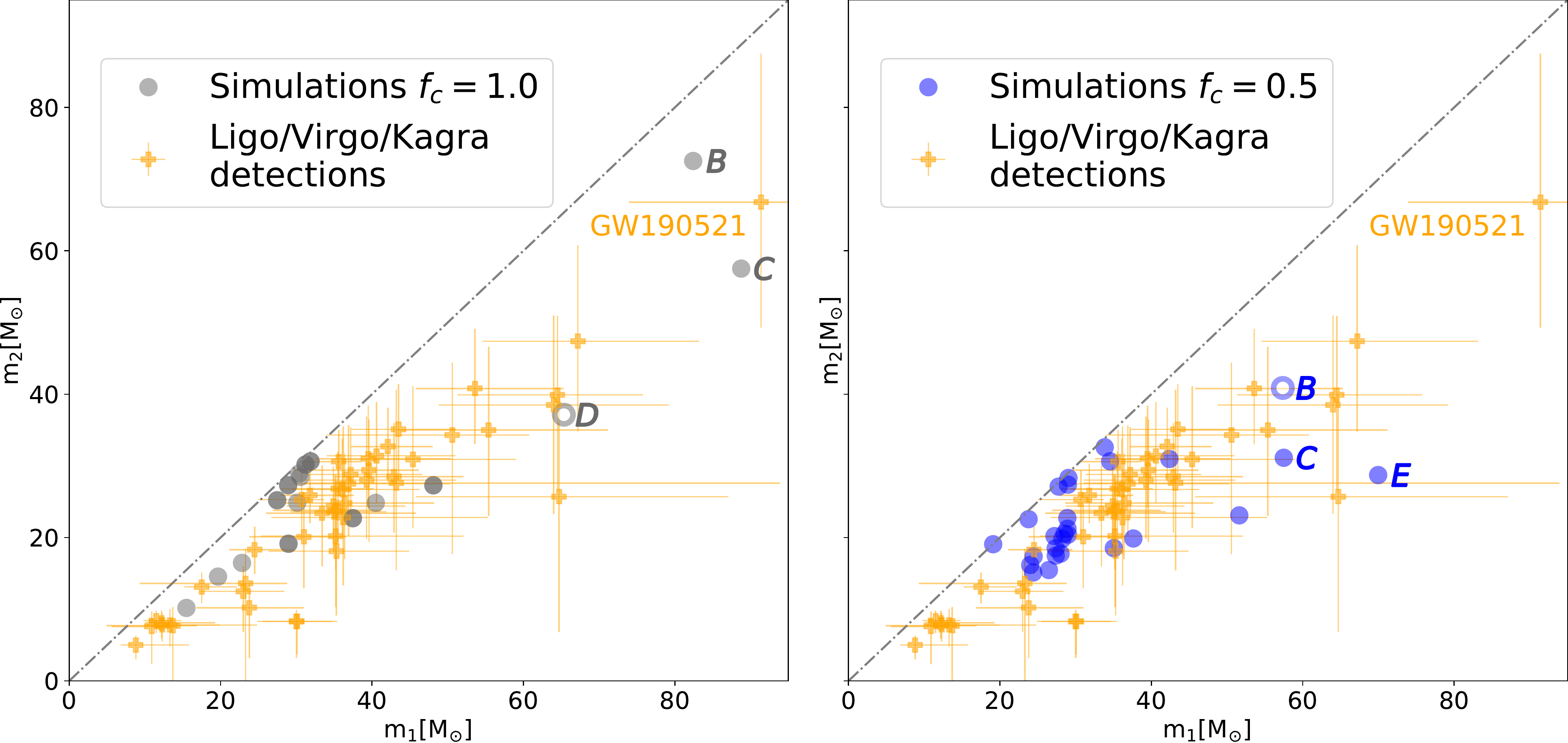}
    \caption{The panels show the primary ($m_1$) and secondary ($m_2$) BH masses of all BH mergers in the simulations for an accretion fraction of \fb (left, grey circles) and \fc (right, blue circles). The letters indicate the simulated formation paths highlighted in Fig. \ref{fig:collision_trees_fc1} and \ref{fig:collision_trees_fc05}. The currently available LIGO/Virgo/Karga gravitational wave detections including error bars are indicated in orange. BH merger events that might be excluded due to gravitational recoil kicks are indicated with open circles. In general, the simulated events cover a similar parameter space as all currently available observations. The \fc simulations provide two possible formation paths for GW190521. Path C is a second-generation event and has a low probability due to a first generation BH merger. Path B is more likely as the event itself is a first generation BH merger.  }
    \label{fig:bh_bh_collisions}
\end{figure*}

Our runs generated a large variety of  BH mergers. The \fc simulations, for example,  resulted in 27 BH - BH collisions (just $2$ more than the \fb runs). These BH merger events cover almost the entire mass range of the LIGO/Virgo/Karga detections as shown in  Fig. \ref{fig:bh_bh_collisions}. \citet{2020MNRAS.498..495D} presented comparable results for a larger set of simulations of lower mass metal-poor star clusters. 
Interestingly, two-star cluster realisations with \fc have produced events with primary and secondary masses very similar to the GW190521 mass-gap detection. These BH merger events are highlighted in panels (B) and (C) of Fig. \ref{fig:collision_trees_fc1}. Path B is particularly interesting as it indicates that BH mergers in the GW190521 mass range could originate from BHs which have grown by BH - star collisions. This formation path is not influenced by BH merger kicks. Therefore GW190521-like events might naturally occur in compact star clusters and they do not necessarily require very massive environments such as nuclear star clusters. The only requirements are that the environment is compact enough to trigger several BH - star mergers and that a large fraction of the stellar mass is accreted onto the BH during the collision.

As we have mentioned above, the simulations presented here do not include relativistic velocity recoil kicks for compact object merger remnants. Therefore all massive BHs formed through two or more BH - BH collisions would have a much lower formation probability in the presence of kicks. The probability to retain a compact merger remnant in the cluster depends on its escape velocity as well as the spin and the mass ratio of the colliding BHs. Following \citet{ArcaSedda2021} we compute the kick velocity distributions for different assumed natal spins and quantify the retention probability of the BH merger remnants. We test three different initial spin distributions for stellar-mass BH, centered around $S_1 = 0.1$ with a Gaussian standard deviation of $\sigma=0.1$, and fixed spins of $S_2 = 0.01$ and $S_3 = 0.001$, respectively.

We compute the retention probability for each BH - BH collision reported in Fig. \ref{fig:collision_trees_fc1} and \ref{fig:collision_trees_fc05} for all three spin distribution and present the results in Tab. \ref{table:spins}.

Under the assumption that stellar BHs are born with very low spins \citep[$S_2$ or $S_3$,][]{Fuller2019}, only the events (B) in Fig. \ref{fig:collision_trees_fc1} and (C) in Fig. \ref{fig:collision_trees_fc05} would survive the relativistic kick. In these cases, since the mass ratio is close to unity, the recoil velocity is lower than the escape velocity.
For all the other cases  $V_{\text{recoil}} \gtrsim 50$ km/s therefore merger remnants have little to no chance to stay bound to the cluster.  

If we assume the stellar BHs to be slowly rotating with distribution $S_1$ \citep[as suggested by  observations, e.g.][]{Abbott2019},
some BH - BH mergers have a non-negligible probability to stay in the star cluster for certain spin orientations. 
Under this assumption the first BH collision reported in the panel (C) of Fig. \ref{fig:collision_trees_fc1} has a probability of 1.9 \% to stay in the cluster because of comparable masses. Therefore, similar events, which in turn resemble GW190521,  could have been hosted in compact but not very massive star clusters as long as the BHs involved in the collisions are born with a low initial spin similar to $S_2$.

Another interesting event, that has non negligible probability to occur (see Tab. \ref{table:spins}), is illustrated in panel (C) of Fig. \ref{fig:collision_trees_fc05}. The plot  shows the merger of a 57 \msunS and a 41 \msunS BH. Due to the similar masses the BHs this collision might indicate a possible formation channel for detected events like GW190701\_203306 \citep[$m_1 \approx 53.6$ \msun, $m2 \approx 40.7$ \msun][]{Abbott2020precession}. 

To estimate the gravitational recoil velocity we implicitly assume that the spin of a BH is not affected during a collision with a massive star. 
While the BH spins of several low mass X-ray binaries are likely to be influenced by mass transfer from the companion into the BH \citep{Podsiadlowski2003, Fragos2015, Sorensen2017}, all the three BH spins measured in high-mass X-ray binaries appear to have been affected very little by mass accretion \citep{Qin2019}. However, even allowing for larger BH spins for star-BH collisions our results would not change.

Two IMBH binaries form at the centre of two different realizations. The first one, a ($173$ \msunS, $153$ \msunS) BH binary, whose formation paths are shown in panels (E) and (F) of Fig. \ref{fig:collision_trees_fc1}, forms in the second realisation of \fc (see Fig.\ref{fig:3_most_massive_bhs}). Right after formation, the two IMBHs evolve into a hard binary. At the end of the simulation ($t \approx 100$ Myr) this binary has an eccentricity of $0.81$ and a semi-major axis of 1.88 AU, therefore, if evolved in isolation, will merge in about 11.5 Gyr. 
Both IMBHs in this binary formed from the direct collapse of a VMS. According to the stellar model we adopted, a star-forming from multiple stellar mergers might retain a small helium core. This in turn leads to a weaker  (P)PSN allowing the star to collapse into a massive BH or even into an IMBH.  We note that this result is connected to significant uncertainties of stellar evolution and we do not include the effect of rotation in VMSs, which could increase stellar winds mass-loss rate \citep{Maeder2000} and it could affect the final stage of the massive star \cite{Burrows2021}.
The second IMBH binary forms in the sixth realization of model \fb (see right panel of Fig. \ref{fig:3_most_massive_bhs}) and it involves two black holes of $113$ and $99$ \msunS (see panels (A) and (E) of Fig. \ref{fig:collision_trees_fc05}). However, since the $99$ \msunS object is the product of a $70$ and $29$ \msunS BH merger, it is very likely to escape the cluster right after formation (see Tab. \ref{table:spins}). Even if it stays in the cluster this IMBH binary would have an expected merging time in isolation greater than the Hubble time (at the end of the run the binary has eccentricity equal to 0.75 and semi-major axes of 3.25 AU). Nevertheless, the merging time could be considerably reduced by chaotic and hierarchical interactions.




\section{Conclusions}

We study a collisional formation scenario for IMBHs and BHs in the (P)PSN mass gap in young compact star clusters utilizing direct N-body simulations carried out with a version of \textsc{nbody6++gpu} that contains relativistic treatments to include the GW energy loss in compact binaries. We rerun $16$ realizations of the model R06W6 introduced in \cite{Rizzuto2021} with updated stellar evolution recipes.
The upgrades follow the implementations in \cite{Banerjee2020} and include new metallicity-dependent stellar winds prescriptions, recipes for the electron-capture supernovae, and treatments for (P)PSN. The latter creates a mass gap in the BH mass spectrum between $45$ and $195$ \msunS (see Fig. \ref{fig:final_bh_mass}). 

Overall our simulations show that $\sim 10^5$ \msunS compact star clusters have a good chance to generate BHs in the (P)PSN mass gap or even in the IMBH mass range ($\gtrsim 100$) as shown in Fig. \ref{fig:bh_distribution_all}. Together with BH - BH collisions, BH - star collisions seem to play an important role in the formation of massive BHs. We parameterize the fraction of stellar mass absorbed by the BH to be unity (\fc) or 50\% (\fb). The Simulations with \fc and \fb can generate BH merger events with primary and secondary BH masses compatible with nearly the whole mass range of LIGO/Virgo/Kagra gravitational wave detections (see Fig. \ref{fig:bh_bh_collisions}). As we do not follow gravitational recoil, we have, a posterior, estimated, the retention probability for a  BH - BH merger in the star cluster and have identified all unlikely formation paths.

The model assuming efficient accretion (\fc) of stellar material in a merger with a BH generates two BH merger events with primary and secondary masses compatible with the mass-gap merger event GW190521 \citep{Abbott2020precession}.
For one event, both colliding BHs  grow entirely through BH - star mergers (see panel (B) Fig. \ref{fig:collision_trees_fc1}). Therefore, the final merger is independent of gravitational recoil kicks and might not only be expected in high escape velocity systems such as nuclear star clusters. GW190521 like mergers might also form in lower mass compact star clusters at low initial metallicity. The only requirement is that the system is compact enough to trigger many BH - star collisions and that the amount of mass accreted by the BH during the collision is close to unity.

The chain of collisions that led to the second GW event involves two BH - star collisions and one  BH - BH collision before the final merger (see panel (C) of Fig. \ref{fig:collision_trees_fc1}). When these two compact objects merge, the final product should experience a recoil kick which depends on the spin orientation and magnitude of the colliding objects. For low spins, $\sim 0.1$ the probability that the product of this BH - BH collision stays in the cluster is low ($1.9 \%$) but non-zero, therefore this formation path is not excluded.  

Contrary to the \fc model, the $8$ realizations of the model \fb did not form BH mergers in the mass range of GW190521 because \fb simulations tend to form less massive progenitor BHs (see Fig. \ref{fig:3_most_massive_bhs}, and right panel of Fig. \ref{fig:bh_distribution_all}). However, the total number of BHs in the (P)PSN mass gap generated by the two models through BH - star collision is comparable (see Fig. \ref{fig:populate_mass_gap}). Even imposing  \fb, BH - star mergers can bring stellar BHs in the (P)PSN mass range.

In one simulation we even form a hard IMBH binary of 153\msunS and 173 \msunS (see (E) and (F) of Fig. \ref{fig:collision_trees_fc1}) which is expected to merge within a Hubble time. Its formation, however, depends on details of the evolution of very massive stars which are only approximately captured by the stellar evolution models used in this study.



\section*{Acknowledgements}
We thank Albrecht Kamlah, Peter Berczik, and Jarrod Hurley for their cooperation by incorporating the new stellar evolution features into NBODY6++GPU.
TN acknowledges support from the Deutsche Forschungsgemeinschaft (DFG, German Research Foundation) under Germany's Excellence Strategy - EXC-2094 - 390783311 from the DFG Cluster of Excellence "ORIGINS".
The authors gratefully acknowledge the Gauss Centre for Supercomputing (GSC) e.V. (www.gauss-centre.eu) for funding this project by providing computing
time through the John von Neumann Institute for Computing (NIC) on the GCS Supercomputers JUWELS and JUWELS-BOOSTER at Jülich Supercomputing Centre (JSC).
MG was partially supported by the Polish National Science Center (NCN) through the grant UMO-2016/23/B/ST9/02732.
SB acknowledges support from the Deutsche Forschungsgemeinschaft (DFG; German Research Foundation) through the individual research grant "The dynamics of stellar-mass black holes in dense stellar systems and their role in gravitational-wave generation" (BA 4281/6-1; PI: S. Banerjee)
\addcontentsline{toc}{section}{Acknowledgements}


\section*{Data availability}
The data underlying this article will be shared on reasonable request to the corresponding author.




\bibliographystyle{mnras}
\bibliography{references} 


\end{document}